\documentclass[manuscript, screen, natbib=true, nonacm]{acmart}
\usepackage{adjustbox}
\usepackage{tabularx}
\usepackage{float}
\usepackage{amsmath}
\usepackage{natbib} 
\usepackage{breqn}
\usepackage{subfigure}
\usepackage{stackengine}

\usepackage{balance}
\AtBeginDocument{%
  \providecommand\BibTeX{{%
    \normalfont B\kern-0.5em{\scshape i\kern-0.25em b}\kern-0.8em\TeX}}}

\setcopyright{none}
\copyrightyear{2018}
\acmYear{2018}
\acmDOI{XXXXXXX.XXXXXXX}

\acmConference[Conference acronym 'XX]{Make sure to enter the correct
  conference title from your rights confirmation emai}{June 03--05,
  2018}{Woodstock, NY}
\acmBooktitle{Woodstock '18: ACM Symposium on Neural Gaze Detection,
 June 03--05, 2018, Woodstock, NY} 
\acmPrice{15.00}
\acmISBN{978-1-4503-XXXX-X/18/06}

\begin{document}
\newcommand{\ranking}[0]{L}
\newcommand{\allitems}[0]{D}
\newcommand{\allusers}[0]{Q}

\newcommand{\doc}[0]{d}
\newcommand{\docs}[1]{\doc_{#1}}
\newcommand{\user}[0]{q}
\newcommand{\users}[1]{q_{#1}}

\newcommand{\prefix}[1]{\ranking_{\le {#1}}}

\newcommand{\rankingPosition}[0]{\ranking(\doc)}
\newcommand{\relevance}[0]{y(d|q)}
\newcommand{\predictedrelevance}[0]{\hat{y}(d|q)}

\newcommand{\group}[0]{g}
\newcommand{\groups}[0]{\mathcal{G}}
\newcommand{\groupname}[1]{\groups^{#1}}

\newcommand{\progroup}[0]{\groups^+}
\newcommand{\nonprogroup}[0]{\groups^-}

\newcommand{\Protected}[0]{\progroup(\ranking)}
\newcommand{\nonProtected}[0]{\nonprogroup(\ranking)}

\newcommand{\alignmentvec}[0]{\groups(\doc)}
\newcommand{\alignmentmat}[0]{\groups(\ranking)}

\newcommand{\populationEstimator}[0]{\hat{\textbf{p}}}
\newcommand{\probabiltyDist}[0]{\hat p}

\newcommand{\attention}[0]{\mathbf{a}}
\newcommand{\attentionvec}[0]{\attention_{\ranking}(\doc)}
\newcommand{\attentionmat}[0]{\attention_{\ranking}}
\newcommand{\exposure}[0]{\boldsymbol\epsilon}
\newcommand{\groupexposure}[0]{\exposure_{\ranking}}

\newcommand{\stochasticRanking}[0]{\pi}
\newcommand{\targetPolicy}{\tau}

\newcommand{\prefd}[0]{\mathrm{PreF}_\Delta}
\newcommand{\AWRF}[0]{\mathrm{AWRF}_\Delta}

\newcommand{\expectation}[2]{\operatorname{E}_{#1}[#2]}

\newcommand{\contprob}[0]{\alpha}
\newcommand{\slowdecay}[0]{\beta}
\newcommand{\skipprob}[0]{\gamma}

\newcommand{\row}[0]{{\mathrm{row}}}
\newcommand{\iteminrow}[1]{\ranking^{-1}(\mathrm{k}, \cdot)}
\newcommand{\iteminrowcol}[2]{\ranking^{-1}(\mathrm{#1}, \mathrm{#2})}
\newcommand{\itemrow}[0]{\row(\doc)}
\title{Towards Measuring Fairness in Grid Layout in Recommender Systems}

\author{Amifa Raj}
\affiliation{%
  \institution{People and Information Research Team\\ Boise State University}
  \city{Boise}
  \state{Idaho}
  \country{USA}
  \postcode{83725-2055}
}
\email{amifaraj@u.boisestate.edu}


\author{Michael D. Ekstrand}
\affiliation{%
  \institution{People and Information Research Team\\ Boise State University}
  \city{Boise}
  \state{Idaho}
  \country{USA}
  \postcode{83725-2055}
}
\email{michaelekstrand@boisestate.edu}


\begin{abstract}

There has been significant research in the last five years on ensuring the providers of items in a recommender system are treated fairly, particularly in terms of the exposure the system provides to their work through its results. However, the metrics developed to date have all been designed and tested for linear ranked lists.
It is unknown whether and how  existing fair ranking metrics for linear layouts can be applied to grid-based displays.
Moreover, depending on the device (phone, tab, or laptop) users use to interact with systems, column size is adjusted using column reduction approaches in a grid-view. 
The visibility or exposure of recommended items in grid layouts varies based on column sizes and column reduction approaches as well.
In this paper, we extend existing fair ranking concepts and metrics to study provider-side group fairness in grid layouts, present an analysis of the behavior of these grid adaptations of fair ranking metrics, and study how their behavior changes across different grid ranking layout designs and geometries. 
We examine how fairness scores change with different ranking layouts to yield insights into (1) the consistency of fair ranking measurements across layouts; (2) whether rankings optimized for fairness in a linear ranking remain fair when the results are displayed in a grid; and (3) the impact of column reduction approaches to support different device geometries on fairness measurement.
%
This work highlights the need to use layout-specific user attention models when measuring fairness of rankings, and provide practitioners with a first set of insights on what to expect when translating existing fair ranking metrics to the grid layouts in wide use today.


\end{abstract}



\maketitle
\section{Introduction}
\label{sec:into}
\begin{figure}[ht]
    
    \subfigure[\label{fig:vertical search} Linear Vertical Layout]{
        \includegraphics[width=0.04\linewidth]{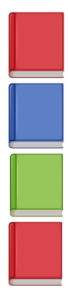}
    }
    \subfigure[\label{fig:horizontal search}Linear Horizontal Layout]{
        \includegraphics[width=0.2\linewidth]{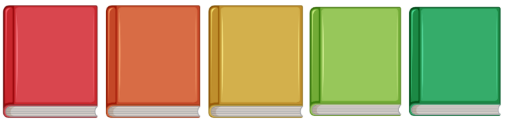}
    }
    \subfigure[\label{fig:wrapped grid rec} Wrapped Grid Layout]{
        \includegraphics[width=0.2\linewidth, scale=0.4]{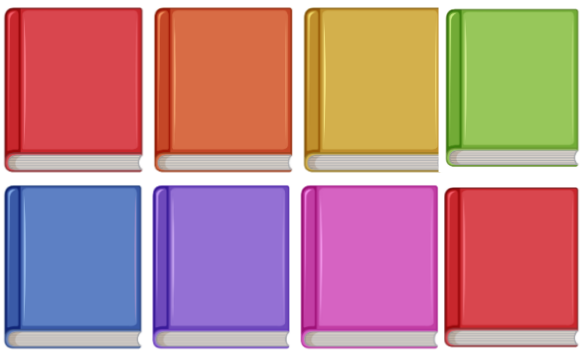}
    }
    \subfigure[\label{fig:LL rec} Multi-list Grid Layout]{
        \includegraphics[width=0.2\linewidth]{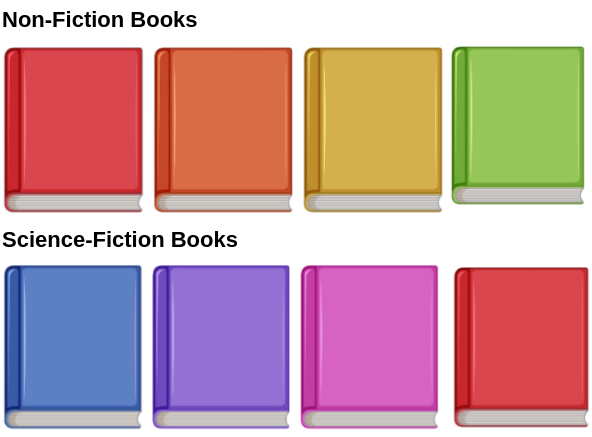}
    }
\caption{Various Types of Layouts}
\label{fig:ranking model}
\end{figure}

Recommender systems may induce unfair distribution of exposure across items and their providers on either individual or group basis, often reflecting societal or historical bias such as prioritizing items from certain races or genders. An ``equality of opportunity'' goal \citep{raj2022measuring} ensures that two providers whose items are equally useful to a user's information need have the same opportunity to be exposed to users, but systems do not always meet this criteria, instead providing \textit{disparate exposure} \cite{diaz2020evaluating}. There are several metrics to measure fairness of exposure (or related constructs) in ranked lists \cite{raj2022measuring}, but they are designed for linear --- usually vertical --- layouts. However, many systems use other ranking layouts such visual grids or voice responses.
Grid layouts (figure~\ref{fig:wrapped grid rec}) are particularly popular for streaming media platforms and image search, but also appear elsewhere; there has been little work to determine how to measure group fairness in such layouts, or how to measure fairness when the system may use different layouts in different contexts or device.
It can be problematic to measure grid-layout fairness by simply mapping the grid positions to a linear layout and using existing metrics, because user attention to items in different positions varies between layouts \cite{chen2022reinforcement}.
For the same set of recommended items, user attention varies depending on how the items are being displayed, affecting item exposure and therefore the fairness of that exposure.
Using fair ranking metrics without taking layout-specific user browsing behaviour into consideration may provide unreliable and erroneous results.

Further, based on the device (phone, tablet, TV, laptop, etc.) used to interact with a system, the geometry of grid layouts varies, often re-ranking the list as the number of available columns changes. 
There are also multiple methods for adjusting the layout: for example, when moving from a wider to a narrower screen, some systems \emph{truncate} the list at the right-side while others \emph{re-wrap} the entire list. 
The impact of these layout adjustments on fairness scores is unknown.
In summary, researchers and developers using grid layouts have little to work with when trying to reason about how the system layouts affect equity of exposure or how to apply the various metrics that have been developed to this setting.
In this paper, we seek to fill this gap and broaden the applicability of fair ranking metric research by extending fair ranking metrics to grid layouts, providing the first (to our knowledge) study of metrics for this widely-used but under-studied paradigm. We observe what happens to group fairness for a list of recommended items with the change of layouts by answering the following research questions:

\textbf{RQ1.} Do fairness measurements remain consistent across layouts?

\textbf{RQ2.} Do rankings optimized for fairness in linear layouts remain fair in grids?

\textbf{RQ3.} How do provider-side group fairness scores change as grid size changes?
    
        \textbf{RQ3.a.} Does the fair ranking metric score change when the grid layout is truncated or re-wrapped?
        
        
        \textbf{RQ3.b} Does the change in fairness score with column-size reduction remain consistent across reduction approaches?

The main contributions of this work are to:
\begin{itemize}
    \item Describe various types of layouts that are often used to display recommended items.
    \item Incorporate grid browsing models into fair ranking metrics to derive fair grid metrics
    \item Provide insights on fairness score consistency and applicability across layouts. 
    \item Describe the impact of column reduction approaches on fairness scores within a grid layout.
\end{itemize}


\section{Background and Related Work}
\label{sec:brw}
This work draws from a line of work on fair RS that we review here and browsing models in sec~\ref{sec:bm}.
Recommender systems often present recommended results in top-$N$ ranked order based on relevance to user information preference. Thus systems \textit{expose} recommended items along with their providers through ranked lists and these ranked lists can be represented in \textit{linear} (figure~\ref{fig:vertical search}) or \textit{grid} (figure~\ref{fig:wrapped grid rec}) layouts.
It is not possible for items with the same relevance to get the same position in a single ranked lists, and a small change of relevance causes item position to vary \cite{singh2018fairness}, thus affecting user attention they receive. RS can cause \emph{disparate exposure} based on provider group association while distributing exposure across relevant items. User attention is not uniformly distributed across items in a ranked list and users tend to interact more with items at the top positions, which causes both economic and reputational disadvantage to the items (and their providers) at lower-ranked positions. Thus, items with similar merit will not receive similar benefits due the position difference or \emph{disparate exposure}. 

The focus of this work is on \textbf{provider-side group fairness} in RS ranking ensuring that different groups of item providers do not experience unjustified discrepancies in the exposure of their content on the basis of their sensitive attributes, such as gender or ethnicity.
Several metrics have been proposed to measure provider-side group fairness in ranking. The broader goal of these metrics is to measure system's ability to allocate fair exposure across item providers based on their group membership and thus they measure exposure discrepancy across groups in ranking. \citet{yang2017measuring, zehlike2022fair}, and \citet{sapiezynski2019quantifying}, among others proposed metrics that measure group fairness for providers in a single ranking. Without considering relevance information these metrics measure fairness as \textit{statistical parity} where item position should not be affected by group membership. Among these metrics \citet{sapiezynski2019quantifying} use \textit{position weights} derived from a user attention model to measure the fairness of item exposure. \citet{singh2018fairness, biega2018equity}, and \citet{diaz2020evaluating} proposed metrics that considered relevance information in fairness measurement and measure fairness as \textit{equal opportunity} where exposure or attention should be proportional to relevance. These metrics measure fairness in sequences or distributions over rankings since it is not possible to achieve fair exposure in a single ranking. All these metrics also consider position weight in their fairness measurement. \citet{beutel2019fairness} and \citet{narasimhan2020pairwise} took a different approach and measure fairness by considering pairwise ordering. 

\citet{kuhlman2021measuring} provided a comparative analysis of selected metrics that can measure statistical parity and \citet{raj2022measuring} provided a comprehensive and comparative analysis of existing metrics that are suitable to measure provider-side group fairness in ranking showing the conceptual similarities and differences among the metrics. 
\citet{raj2022measuring} identified metrics that considers user attention changing behaviour with item positions to determine position weight in ranking. Various user browsing models are considered to demonstrate user browsing behaviors in ranking and their study showed that metrics show sensitivity towards the choice of user browsing models.




\section{Problem Formulation}
\label{sec:problem-grid}
\begin{table}[tb]
\centering
 \footnotesize
 \caption{Summary of notation.}
 \begin{tabular}{cl}
 \toprule
  $\doc \in \allitems$ & document or item \\
  $\user \in \allusers$ & request (user or context) \\
  $\ranking$ & ranked results of $N$ documents from $\allitems$ \\
  $\ranking^{-1}(i)$ & the document in position $i$ of linear (1-column) layout \\
  $\rankingPosition$ & rank of document $d$ in linear layout \\
  $\itemrow$ & row number of document $d$ in grid layout \\
  $\iteminrow{k}$ & items in $k$th row in grid layout \\
  $\iteminrowcol{k}{c}$ & items in row $k$ and column $c$ in grid layout \\
  $\relevance$ & relevance of $d$ to $q$ \\
  $\group$ & number of groups \\
  $\alignmentvec$ & group alignment vector \\
  $\alignmentmat$ & group alignment matrix for documents in $\ranking$ \\
  $\Protected$ & set of documents in protected group in $L$ \\
  $\nonProtected$ & set of documents non-protected group in $L$ \\
  $\populationEstimator$ & target group distribution \\
  $\attentionmat$ & attention vector for documents in $\ranking$ \\
  $\attentionvec$ & position weight of $\doc$ in $\ranking$ \\
  $\groupexposure$ & the exposure of groups in $\ranking$ \\

\bottomrule
 \end{tabular}
 \label{tab:notation}
\end{table}

In this work, we consider a recommender system that recommends $n$ items $\docs{1}, \docs{2}, \dots, \docs{n}$ $\in \allitems$ in response to information requests from users $\users{1}, \users{2}, \dots, \users{m} \in \allusers$ based on their relevance to the request $\relevance$ and presents the results in a layout $\ranking$ (either 1-column, as in a classical linear layout, or a multi-column layout). Documents are associated with producers or providers who in turn can be associated with demographic attributes identifying them with one or more of $\group$ 
groups.
We model group membership of documents with group alignment vector $\alignmentvec \in [0,1]^g$ (s.t. $\|\alignmentvec\|_1=1$) forming a distribution over groups; this allows for mixed, partial, or uncertain membership in an arbitrary number of groups.
Table~\ref{tab:notation} summarizes the notation used in this paper. 

\subsection{Ranking Layouts} 
\label{sec:layout}

Without loss of generality, we treat recommendation and layout as a multi-stage process: the system first scores and ranks items for the user (either deterministically or with a stochastic policy \citep{diaz2020evaluating}), and then displays that ranking in a layout. In this work, we consider layouts in $r \times c$ grids, where $r$ is the number of rows and $c$ the number of columns; this encapsulates at least four distinct models.
One such family of layouts comprise the familiar \textbf{linear layouts} where items are displayed in a single linear list. These come in two varieties
\begin{description}
    \item[Vertical Ranking Model] Items are displayed in a multi-row single-column ranked list ($r\times1$, see Figure~\ref{fig:vertical search}). Users generally see items from top to bottom.  The layout may be split into multiple pages.
    \item[Horizontal Ranking Model] Items are displayed in single-rows with multiple column lists following $1\times c$ pattern, see figure~\ref{fig:horizontal search}. Users see items from left to right.
\end{description}
In \textbf{grid layout}, items are displayed in multiple rows and columns ($r \times c$).  These also come in multiple varieties:
\begin{description}
    \item[Wrapped Grid] Items are displayed as a single ranking in an $r \times c$ grid, without being categorized into groups, see figure~\ref{fig:wrapped grid rec}.
The grid is formed by displaying the items in order horizontally and starting a new row when the display runs out of space.
    \item[Multi-ranking Grid]Items are displayed in multiple rows, often based on categories or recommendation sources, and each row consists of a ranked list of items.
In figure~\ref{fig:LL rec}, recommended items are categorized by genre which may facilitate users to find them from their preferred categories.
\end{description}

We focus on \textbf{wrapped grid} layout in this work due to the better availability of browsing and attention models for this problem setting.
Further work is needed to provide usable models of user browsing behavior with multi-ranking grids before we can attempt to measure their fairness.
\subsection{Fair Ranking Metrics}
We follow the recommendations of \citet{raj2022measuring} and study two metrics: \textit{Attention-Weighted Rank Fairness} \citep[$\AWRF$,][]{sapiezynski2019quantifying} to measure \textit{statistical parity} in single ranking (averaging over multiple rankings to measure overall system fairness), and \textit{Expected Exposure Loss} \citep[$\mathrm{EEL}$,][]{diaz2020evaluating} to measure equal opportunity in sequences of rankings. These metrics measure the distribution of exposure (based on estimated user attention) across provider groups to measure the fairness of rankings. They represent user attention with a \textit{position weight} assigned to each document in a ranking. 

Both metrics rely on a model of user attention (estimating the attention a user is likely to give to items at different positions in ranking) in order to measure fairness; it is important to know how users browse and interact with different positions in the ranked layout.
Several studies have used user eye gaze tracker to study user browsing behavior \cite{shrestha2007eye, djamasbi2011visual, xie2017investigating, zhao2016gaze}. Some studies used user click behavior to infer browsing behavior of users \cite{xie2017investigating, dupret2008user} with respect to ranking positions.
Simple models of user browsing behavior, commonly used in information retrieval metrics and described in the next sections, determine these weights based on items' position in the ranking (along with other information, such as the relevance of preceding documents).

$\AWRF$ is suitable to measure provider-side fairness in single ranking and it measures the difference between group exposure and configurable \emph{population estimator} (representing the ideal distribution of exposure over groups) using a distance function $\Delta$. The exposure for each groups $\groupexposure$ is derived from the attention vector and the group alignment matrix ($\mathrm{\groupexposure = \alignmentmat^T \attentionmat}$) which aggregates the attention given to items of each group in proportion to their group membership as represented by the alignment vector: 
\begin{equation}
\label{eq:AWRF}
    \AWRF(\ranking) = \Delta(\groupexposure, \populationEstimator) 
\end{equation}


$\mathrm{EEL}$ is suitable for \textit{stochastic} ranking policy where fairness is measured over user-dependant distribution over rankings $\rho(\ranking|\user)$ since it is not possible to achieve equal exposure in single ranking \cite{diaz2020evaluating}. It can be drawn as distribution over rankings $\ranking_1,\ranking_2,\dots,\ranking_{\tilde n}$ from the distribution over requests $\rho(\user)\stochasticRanking(\ranking|\user)$ \cite{raj2022measuring}.
$\mathrm{EEL}$ uses available relevance information to derive a \emph{target exposure} $\exposure_\targetPolicy$, based on an ideal policy $\targetPolicy$ where relevant items are sorted in non-decreasing order in ranking and exposure is fairly distributed across the relevant items.  Using the $\groupexposure$ of each ranking, the system exposure is derived as $\exposure_\stochasticRanking = \sum_L \stochasticRanking(\ranking|\user)\exposure_\ranking$. $\mathrm{EEL}$ is computed as the squared Euclidean distance between system exposure $\exposure_\stochasticRanking$ and target exposure $\exposure_\targetPolicy$: 
\begin{align}
    \mathrm{EEL} & = \|\exposure_\stochasticRanking - \exposure_\targetPolicy\|_2^2 
     \label{eq:eel-decomp}
\end{align}

Since these metrics compute fairness as a distance from the target distribution regardless of layout, it is meaningful to directly compare fairness scores between layouts for the same test data.


\subsection{Linear Browsing Models}
In linear ranking, users typically browse the list from top to bottom \cite{craswell2008experimental}, with the probability that they will continue (and thus view more items) decreasing as they move down the list. 
There are a variety of models of this scanning behavior with decaying attention; \textit{cascade} and \textit{geometric} are commonly-used click models to estimate user interaction probability with ranking positions. These models have been employed to construct evaluation metrics to measure utility \cite{moffat2008rank, carterette2011system, chapelle2009expected, ashkan2011informativeness} or item exposure \cite{diaz2020evaluating, biega2018equity, sapiezynski2019quantifying} in rankings.

\citet{moffat2008rank} proposed the \textit{rank-biased precision} (RBP) evaluation metric to weight precision based on user attention to different ranking positions. This metric used a geometric browsing model with a \textit{continuation probability} $\contprob$ to estimate the probability of users moving to the next item (position) or stopping (click) at that position; the visiting probability exponentially decreases with ranking positions. \citet{biega2018equity} proposed a modified version where the position weight decays geometrically with each position having the same probability of being stopped (clicked). 
In this model, the visiting probability of item $d$ in position $\rankingPosition$ is determined by:
\begin{equation}
    P_{\mathrm{geometric}}[V_d] = \contprob^\rankingPosition
\end{equation}
\citet{craswell2008experimental} proposed the \textit{cascade} click model where users will view position $i$ if they have skipped items before that position, and whether users will click or skip a position depends on the relevance of the item in that position and the relevance of items in previous positions. \citet{chapelle2009expected} proposed a cascade-based metric \textit{expected reciprocal rank} (ERR) by extending the cascade model to include the probability of users terminating the entire process as an \textit{abandonment} probability that decays geometrically. 
In the cascade model, users will visit item $d$ if they did not stop at any position before that item in the ranked list which is determined by item relevance.
The continuation probability $\contprob$ is now a function of relevance, and the probability of visiting $d$ is given by:
\begin{equation}
    P_{\mathrm{cascade}}[V_d] = \prod_{j\in[0,\rankingPosition)}\contprob \left(y\left(\ranking^{-1}\left(j\right)\middle|q\right)\right)
\end{equation}



\begin{table}[]
\caption{Parameters of Weighting Models for computing $\attentionvec$ and the range of parameter values }
\label{tab:grid_browse}
\footnotesize
\begin{tabular}{llll}
\textbf{Parameters}                  & \textbf{Values}        & \textbf{Browsing Models}       & \textbf{Default Values}                              \\ \hline
Skipping Probability $\skipprob$     & \{0.1, 0.2, ..., 0.9\} & Row Skipping        & 0.5                                        \\ \midrule
Continuation Probability $\contprob$ & \{0.1, 0.2, ..., 0.9\} & \begin{tabular}[c]{@{}l@{}}Cascade\\ Geometric\end{tabular}  & 0.5\\ \midrule
Slow parameter $\slowdecay$                         & \{1.1, 1.2, ..., 2.0\} & Slower Decay     & 1.9                                           \\ \hline
\end{tabular}
\end{table}

\subsection{Grid-based Browsing Models}
\label{sec:bm}
Users do not interact with grid displays the same way they interact with linear displays and several studies have been performed to understand how users allocate attention to different items in grid layouts.

\subsubsection{Existing Literature on User Browsing Behavior in Grid Layouts}
\citet{tatler2007central} observed that users show tendency of \textit{central fixation} where they tend to put more attention on the middle of the screen than on the edges, \citet{djamasbi2011visual} found that users usually focus on results located at the top left-hand side and proceed in an \textit{F-shaped} reading pattern, but the viewing pattern varies depending on task, content, and complexity of web pages. The eye-tracking study of \citet{zhao2016gaze} also observed an \textit{F-pattern} in user interaction with grid-based recommendations but the pattern can vary depending on task
\citet{shrestha2007eye} emphasized on the need of considering page content while understanding user viewing patterns.
\citet{xie2019grid, xie2017investigating} performed eye-tracking studies in grid-based image search results and observed the \textit{middle bias} pattern. Moreover, in grid-view, user attention decreases at a slower rate than in linear layouts (\textit{slower decay}) and users often jumps to results after skipping rows (\textit{row skipping}).

The studies mentioned above mostly focus on understanding user viewing patterns in grid-based interfaces with the goal of providing and measuring user satisfaction. There is limited research work concerning fairness issues when results are displayed in grid layout.
\citet{guo2020debiasing} proposed de-biasing techniques for grid-based product search result pages in e-commerce systems; consistent with the studies above, they observed that user attention follows \textit{row skipping} and \textit{slower decay} while interacting results in grid layout. \citet{balyan2021using} emphasized on item meta information in user viewing behavior in grid-based e-commerce search results.
\subsubsection{Adapting Browsing Models to Grid-Based User Behavior}
Since the previous studies showed that user attention varies between applications depending on task, domain, device, and details of the layout, considering multiple viable models from existing literature will provide insights useful to researchers and practitioners in various contexts, as they can apply an appropriate model for their systems. For this preliminary analysis, we will consider \emph{row-skipping} (RS) and \emph{slower-decay} (SD) in the context of wrapped grid layout; we leave central fixation, multi-list rankings, and incorporating multiple browsing model adjustments simultaneously to future work.

We adapt both the \emph{cascade} and \emph{geometric} browsing models to account for \emph{row-skipping} (RS) and \emph{slower-decay} (SD). Table~\ref{tab:grid_browse} shows the parameters and range of values we consider to measure attention weight of items in ranking.
For \emph{row skipping} behavior, the visiting probability of item $d$ at $\itemrow$ and ranking position $\rankingPosition$ depends on the skipping probability of a row $\skipprob$; for each of the $k$ rows before $\itemrow$, the user either continued through that row, or skipped it with probability $\skipprob$. If user visited items in a row, that implies that a particular row was not skipped. 
With that assumption, visiting probability of item $d$ in cascade-based row-skipping model considering relevance:
\begin{equation}
    P_{RS(\mathrm{cascade})}[V_d] = \left[\prod_{k=0}^{\itemrow}(1-\skipprob) \prod_{i\in \iteminrow{k}} \contprob (y(\ranking^{-1}(i)|\user)) + \prod_{k=0}^{\itemrow}\skipprob\right] \prod_{i\in \itemrow}\contprob (y(\ranking^{-1}(i)|q))
\end{equation}

The visiting probability of item $d$ in geometric-based row-skipping model is given by\footnote{This model is derived by \cite{guo2020debiasing}) where they referred the model as \textit{cascade click model}. However, in our paper, we referred the model as \textit{geometric} to keep the conceptual consistency.}:
\begin{equation}
    P_{RS(\mathrm{geometric})}[V_d] = \left[\prod_{k=0}^{\itemrow}(1-\skipprob) \prod_{i\in \iteminrow{k}} \contprob + \prod_{k=0}^{\itemrow}\skipprob\right] \prod_{i\in \itemrow}\contprob
\end{equation}

With the \emph{slower-decay} browsing behavior, visiting probability of items across a row in a grid layout decays more slowly than in a vertical linear list, but jumps when the user moves to the next row.
This is modeled by a decay parameter $\slowdecay$ to modify the continuation probability for items in ranked results based on the row in which they appear.
The visiting probability of item $d$ in cascade-based slow-decay model is:
\begin{equation}
    P_{SD(\mathrm{cascade})}[V_d] = \operatorname{min}(\slowdecay^{\itemrow} \prod_{i=[0, \rankingPosition]}\contprob (y(\ranking^{-1}(i)|q)), 1)
\end{equation}
The geometric visiting probability of item $d$ with slower decay is (derived by \cite{guo2020debiasing}):
\begin{equation}
    P_{SD(\mathrm{geometric})}[V_d] = \operatorname{min}(\slowdecay^{\itemrow} \prod_{i=[0, \rankingPosition]}\contprob, 1)
\end{equation}

\subsection{Changing Grid Layouts}
\label{sec:col-red}
Based on the device the users use to interact with the system, grid layout can be converted into a size suitable for a particular device using two different approaches: \emph{truncation}, where each row is truncated and item off-screen are no longer displayed, and \emph{re-wrapping}, where the rows are re-wrapped so the items that would be off-screen are moved to the next row. These approaches may differ in their influence on the fairness of the resulting display. 
To observe the impact of column size and column reduction approaches on group fairness score, we change the column size for a given grid ranking using both \emph{truncation} and \emph{re-wrap} approaches.

\section{Experimental Setup}
\label{sec:exp}
Our central goal is to understand how measurements and optimizations for classical linear rankings apply to grid layouts, both to apply existing methods and to identify where further research is needed to support fairness in these widely-used layouts.
To answer our research questions, we conduct several experiments by implementing the metrics with adaptations for user behavior in grid layouts and using them to measure outputs in real-world datasets.

\subsection{Dataset}
We use two user-book interaction datasets from \textit{GoodReads} \cite{wan2018item} and \textit{Amazon} \cite{mcauley2015image}, integrated with the PIReT Book Data Tools\footnote{\url{https://bookdata.piret.info}} \cite{ekstrand:bag} to obtain author metadata. Table~\ref{tab:data} shows the summary of the datasets. For both datasets, we generate 1000 personalized book recommendations for 5000 users using four collaborative filtering algorithms: user-based (UU \cite{10.1145/312624.312682}), item-based (II \cite{deshpande2004item}), matrix factorization (WRLS \citep{Takacs2011-ix}), and Bayesian Personalized Ranking (BPR \cite{rendle2012bpr}), as configured by \citet{ekstrand:bag}. We used \emph{Lenskit for Python} \citep{ekstrand2020lenskit} to generate recommendations using binary implicit feedback for items. 
Author gender identity is the sensitive attributes for our experiments. Due to limitations of the underlying data set \citep{ekstrand2018exploring}, this is a discrete but possibly unknown binary gender attribute; we acknowledge the importance of more faithful representations of gender in research \cite{pinney2023much}, and the metrics that we study can all be used with a larger set of gender identities as well as mixed or partial membership when such data can be obtained.
Both datasets have incomplete relevance judgements and incomplete group labels. We follow common practice and consider documents without relevance data as non-relevant, and treat missing group labels as a separate unknown category in our experiments.
\begin{table}[]
\caption{Summary of experiment data with nDCG score}
    \label{tab:data}
\footnotesize
\begin{tabular}{lccccccccc}
\hline
\multicolumn{1}{c}{\textbf{Dataset}} & \multicolumn{3}{c}{\textbf{Data Statitstics}}               & \multicolumn{2}{c}{\textbf{Group Sizes}} & \multicolumn{4}{c}{\textbf{nDCG}}                        \\ \hline
                                     & \textbf{\#Users} & \textbf{\#Items} & \textbf{\#Test Users} & \textbf{$|\progroup|$}        & \textbf{$|\nonprogroup|$}       & \textbf{II} & \textbf{UU} & \textbf{WRLS} & \textbf{BPR} \\ \hline
Amazon                               & 8,026,324        & 2,268,142        & 5000                  & 217032              & 490953             & 0.08        & 0.13        & 0.10          & 0.03         \\
GoodReads                            & 870,011          & 1,096,636        & 5000                  & 177359              & 282857             & 0.23        & 0.24        & 0.26          & 0.13         \\ \hline
\end{tabular}
\label{tab:my-table}
\end{table}

\subsection{Methodology}

Across several different scenarios, we measure changes in the fairness scores themselves, as well as changes in the ranking of systems (which system is measured to be most fair), to understand the impact of layouts and browsing models on fairness measurement.
\paragraph{\textbf{RQ1}}
To observe the consistency of fair ranking measurements across layouts, we represent the recommended items in linear-vertical layout and 5-column wrapped grid layout and measure fairness using the metrics in their default parameter settings. We implement $\AWRF$ and $EEL$ with the modified user attention models to account for wrapped grid layout. The metric score comparison shows how the fairness scores change with the choice of layouts.

\paragraph{\textbf{RQ2}}
To better understand the fairness score differences across layouts, we identify if fairness remains consistent across layouts --- whether a ranked result optimized to be fair for a certain layout remains fair for other layouts. 
We apply the GreedyEQ group-fair reranking technique from \citep{ekstrand2018exploring} to recommendation results to generate fairness-optimized ranked lists; we then render these rankings into 5-column grid layouts and measure group fairness in both linear and grid layouts.
This experiment shows the persistence of fairness scores of a ranked list across layouts. 
\begin{figure}[ht]
    
    
    
    \subfigure[\label{fig:defualt_rec} GoodReads Recommendations]{
        \includegraphics[width=0.48\columnwidth]{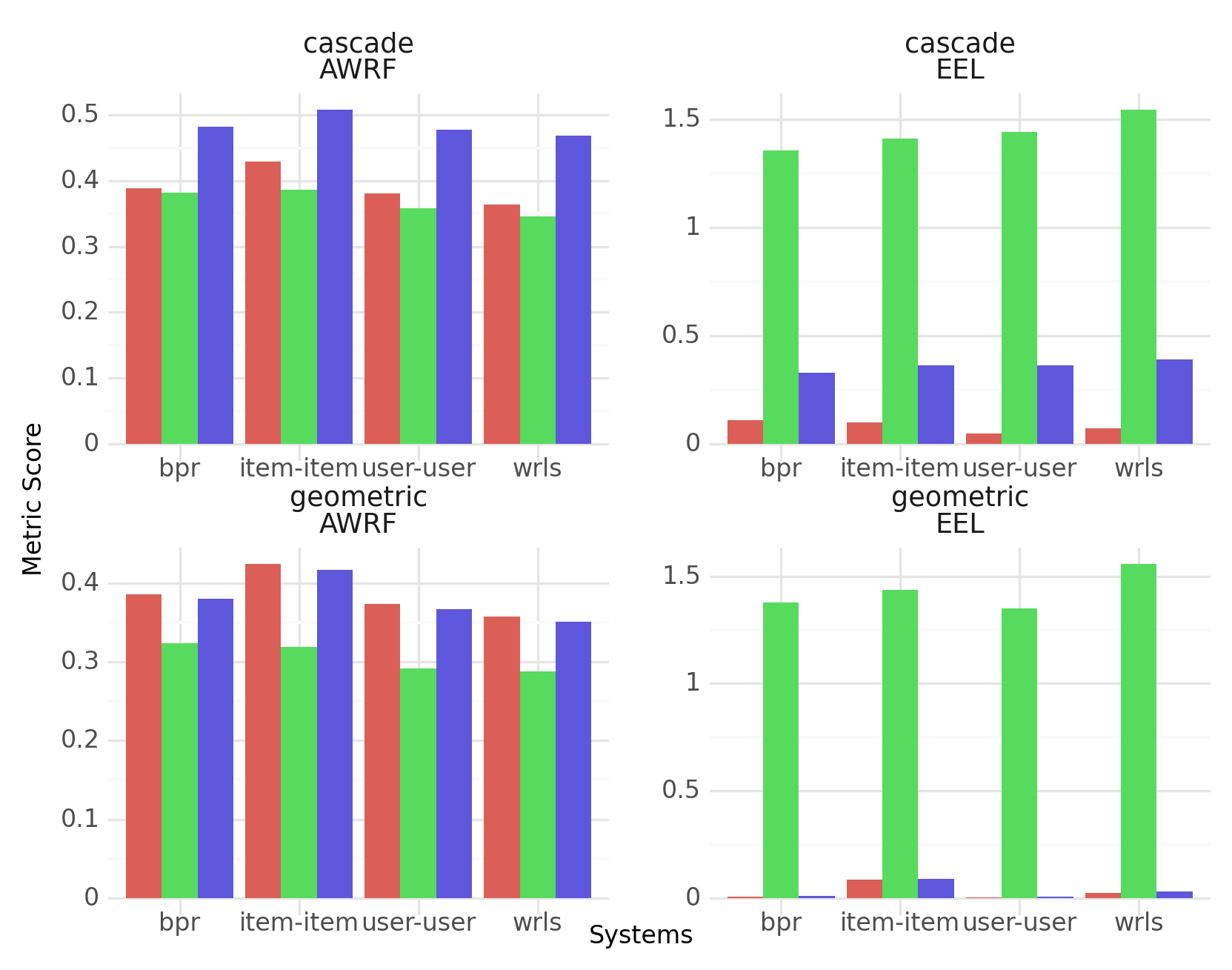}
    }
    \subfigure[\label{fig:rr_rec_opt} GoodReads Fairness-aware Re-Ranked Recommendations]{
        \includegraphics[width=0.48\columnwidth]{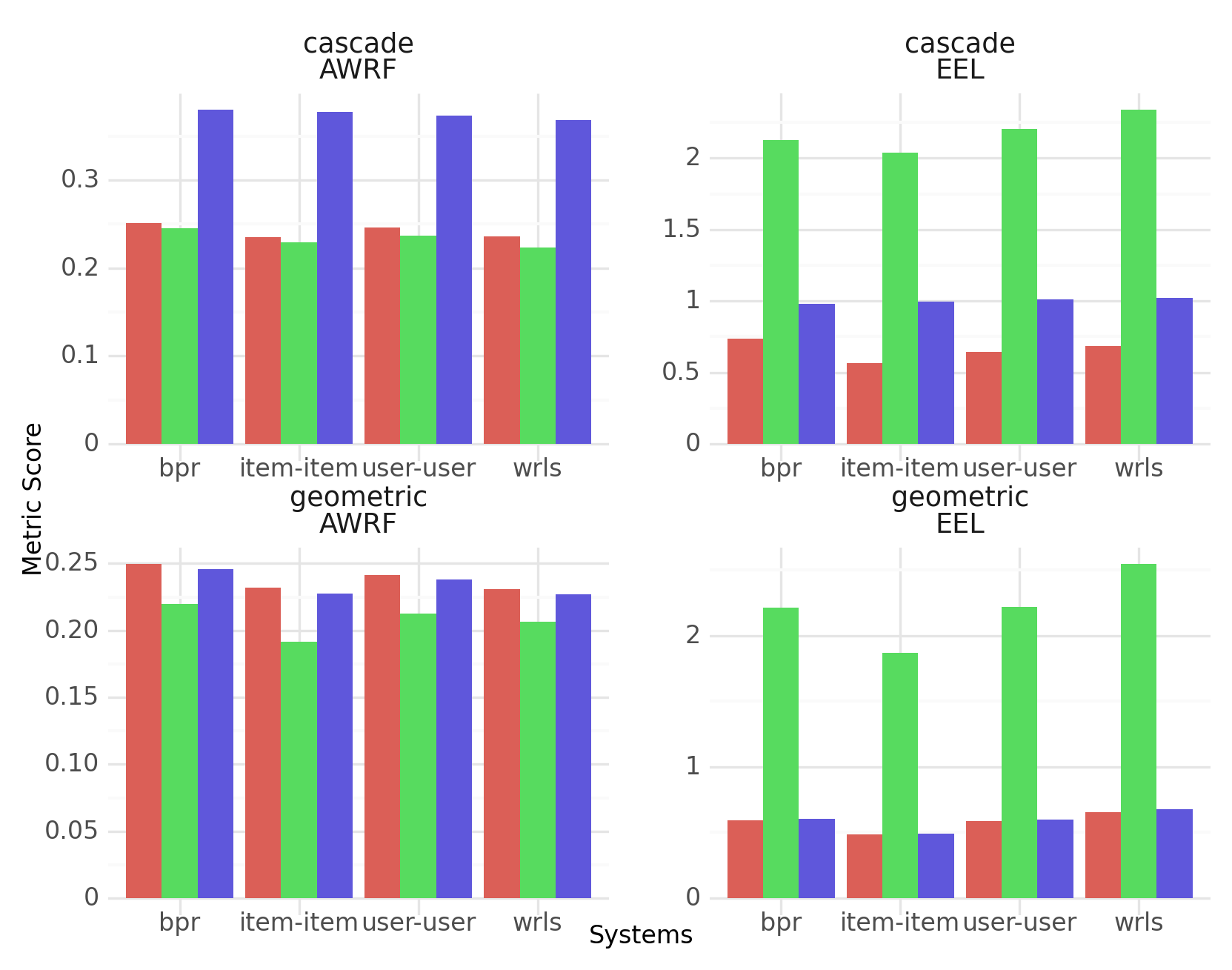}
    }
    \subfigure{
        \includegraphics[width=0.2\columnwidth]{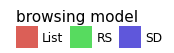}
    }
\caption{Metrics results with the change of weighting strategy}
\label{fig:default}
\end{figure}

    
    
\paragraph{\textbf{RQ3}}

As noted in Section~\ref{sec:col-red}, column reduction can be done by either \emph{truncating} or \emph{re-wrapping} the rows to fit the user's current screen which may have different impacts on the fairness scores of system outputs. Further, fairness scores may change as column size changes regardless of approach. We represent the set of recommended items in grid layout changing the column size in 10, 8, 6, 5, 4, 3 using both truncation and re-wrap approaches. 
To see the impact of column size on group fairness score and the fairness score consistency across column-reduction approaches, we compare fairness scores across column sizes and across the reduction approaches.
\begin{figure}[h]
    
    
   
    \subfigure[\label{fig:cr_awrf} $\AWRF$ in GoodReads Recommendations]{
        \includegraphics[width=0.48\columnwidth]{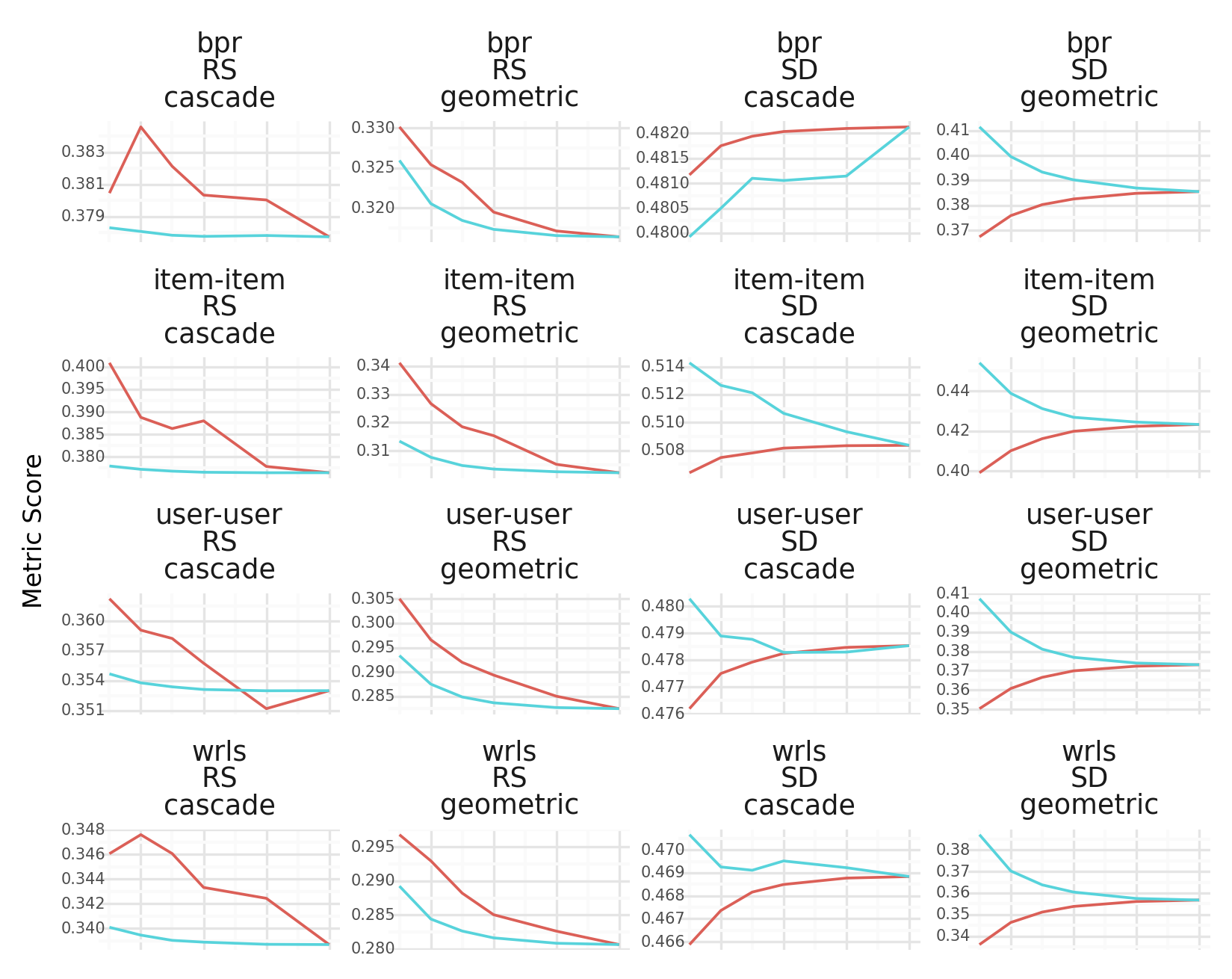}
    }
     \subfigure[\label{fig:cr_ee} $\mathrm{EEL}$ in GoodReads Recommendations]{
        \includegraphics[width=0.48\columnwidth]{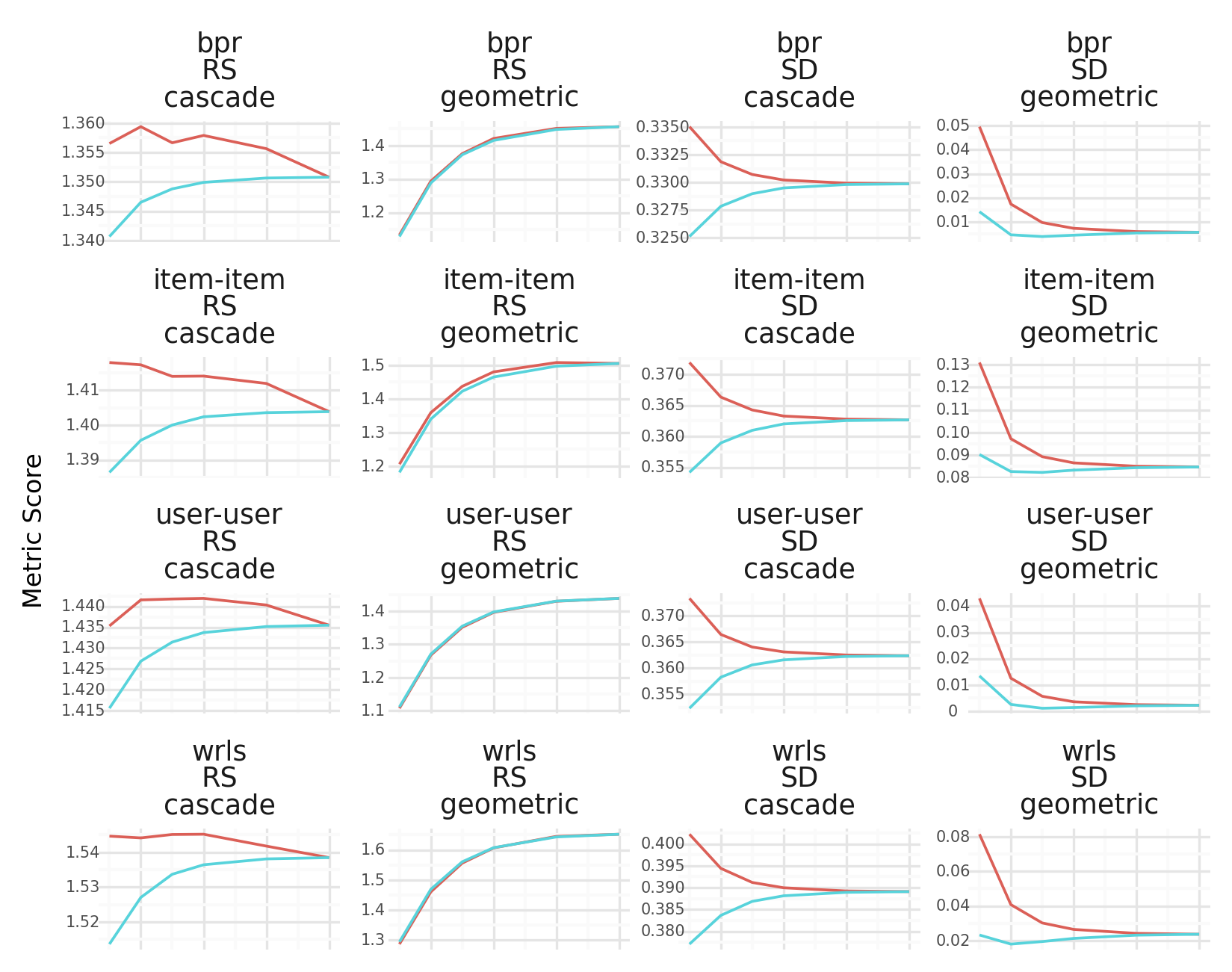}
    }
     \subfigure{
        \includegraphics[width=0.2\columnwidth]{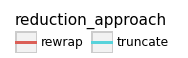}
    }
    
\caption{Metrics results with the change of column sizes across column reduction approaches}
\label{fig:colred}
\end{figure}

    
    
\section{Results and Discussion}

We now present the results of our investigation into the behavior of fair ranking measurements applied to grid layouts. We observe similar results for both datasets and due to space limitations we show results from GoodReads dataset.

\subsection{RQ1: Do fairness measurements remain consistent across layouts?}
Figure~\ref{fig:defualt_rec} shows the fair ranking metric scores change with layouts and within grid layout with the change of browsing behaviors (\textit{row-skipping}, \textit{slower-decay}). $\AWRF$ score varies across grid adjustments to browsing models, keeping the same order of systems, but the \emph{cascade} and \emph{geometric} browsing models rank systems in a different order. $EEL$ scores with \emph{row-skipping} model notable vary; this shift is significantly greater than the shift seen in $\AWRF$.

\textit{Implications.} From \emph{RQ1}, we have following observations: 
\begin{itemize}
    \item Fair raking metric scores are highly dependent on layout and user browsing model.
    \item Within a layout, metric score further varies across user browsing behavior.
    \item Since user attention is one of the required components of $\AWRF$ and $\mathrm{EEL}$ implementation and user attention for ranking positions is determined by user browsing behavior, it is important to consider accurate browsing model while applying these metrics.
\end{itemize}
\subsection{RQ2: Do rankings optimized for fairness in linear layouts remain fair in grids?}
From Figure~\ref{fig:rr_rec_opt}, we see that $\AWRF$ scores are consistent across layouts specifically with \emph{geometric} browsing model. $EEL$ score for a fairness optimized ranking can vary across layouts depending on user browsing models. Within grid layout, $EEL$ with the \emph{row-skipping} browsing model provides different fairness scores and rankings than \emph{slower-decay}. 

\textit{Implications.}
From \emph{RQ2}, we made following observations:
\begin{itemize}
    \item A ranking that is fair in linear layout can be represented as unfair depending on the assumed user browsing behavior. This reinforces the need to incorporate accurate user browsing models in fairness measurement.
    \item Without considering layout-suitable browsing models, metrics will provide unreliable fairness scores. 
\end{itemize}

\subsection{RQ3: How do fairness scores change as grid size changes?}
Figure~\ref{fig:colred} shows how metric score changes with column sizes and the changing pattern with column reduction approaches. 
\paragraph{RQ3.a. Does the fair ranking metric score change when the grid layout is truncated or re-wrapped?}
When columns are reduced using the \textit{truncate} approach, metrics show some stability towards column size for most of the systems. However, column size has more impact on $\AWRF$ scores than $\mathrm{EEL}$ with the \emph{truncate} approach.
When columns are reduced using the \textit{re-wrap} approach, $\AWRF$ shows high sensitivity towards column sizes.
 
\paragraph{RQ3.b Does the change in group-fairness score with column size reduction remain consistent across truncation and re-wrap approach?}
Metric scores vary with the change of column sizes and the direction of this change is different between column reduction approaches. However, for some systems, metric scores with both column reduction approaches converges at some column sizes.
In both datasets, the metrics are consistent across systems.

We do note that the \emph{truncate} approach is primarily used with multi-list layouts in practice, while our results here are for wrapped layouts; however, finding that the use of truncation has significant effects on fairness has implications for fair layouts regardless of the initial grid layout method.

\textit{Implications.} From \emph{RQ3} we have following observations: 
\begin{itemize}
    \item Device is an important factor in measuring fairness.
    \item With the change of device (column size) fairness scores show high sensitivity which indicates the importance of carefully selecting column-reduction approaches while re-ranking the grid layout. 
\end{itemize} 

\subsection{Discussion}
In this work, we consider a gap in the state of the art in measuring the provider-side fairness of rankings by considering grid layouts. We apply existing fair ranking metrics in linear and grid layouts to identify their consistency across layouts.
Our findings provide insights on implementation and reliability of fair ranking metrics in grid layout and provides knowledge on how metric behavior changes across ranking layouts and across column-reduction approaches within grid layouts. Our results suggest that metrics can vary in their consistency across ranking layouts ($\AWRF$ was more consistent across layouts than $EEL$). However, a metric that is consistent across layouts may not be stable across device sizes within a particular grid layout ($EEL$ was more consistent across column sizes, while the consistency of $\AWRF$ metric results notably varies depending on the column-reduction approach.) Therefore, our results advise researchers and practitioners to pay close attention to ranking layout, device sizes, and column-reduction approaches while using a metric to measure fairness in ranking. Even though $\AWRF$ metric score is consistent across layouts to some extent, while using $\AWRF$ in grid layouts, practitioners should pay attention to column sizes and column-reduction approaches, whereas while using $EEL$ to measure fairness in ranking, ranking layout must be taken into account but the reduction approach has less impact on the measurements. 

Furthermore, our results indicate that metrics can be highly affected by user browsing behavior. Since the concept of provider-side fairness in ranking often relies on the attention in different positions, it is important to use accurate models of user attention behavior when measuring provider-side fairness in ranking. It is necessary to develop a clear and detailed understanding of user browsing behavior in order to generate valid and trustworthy fairness score using fair ranking metrics. Our work is not able to directly provide those measurements, but provides a first analysis of what to expect when applying existing measurements with the current public state of knowledge in user behavior modeling.
\bibliographystyle{ACM-Reference-Format}
\bibliography{reference}


\end{document}